\documentclass[preprint]{revtex4}
\usepackage[T1]{fontenc}
\usepackage{graphicx}
\usepackage{rotating}
\usepackage{bm}        
\usepackage{amssymb}   
\usepackage{bm}
\usepackage{float}
\usepackage{tikz}
\usepackage{diagbox}

\def\prb#1#2#3{{\it Phys.~Rev.~B}~{\bf #1},\ #2\ (#3)}

\def\jpcb#1#2#3{{\it J.~Phys.~Chem.~{\rm B}}~{\bf #1},\ #2\ (#3)}
\def\jpca#1#2#3{{\it J.~Phys.~Chem.~{\rm A}}~{\bf #1},\ #2\ (#3)}
\def\jpcc#1#2#3{{\it J.~Phys.~Chem.~{\rm C}}~{\bf #1},\ #2\ (#3)}
\def\jcp#1#2#3{{\it J.~Chem.~Phys.}~{\bf #1},\ #2\ (#3)}

\def\prl#1#2#3{{\it Phys.~Rev.~Lett.}~{\bf #1},\ #2\ (#3)}

\def\mp#1#2#3{{\it Mol. Phys.}~{\bf #1},\ #2\ (#3)}
\def\jpb#1#2#3{{\it J. Phys. B: At. Mol. Opt. Phys.} {\bf #1},\ #2\ (#3)}

\def\pccp#1#2#3{{\it Phys. Chem. Chem. Phys.}~{\bf #1},\ #2\ (#3)}

\def\jctc#1#2#3{{\it J. Chem. Theor. Comp.}~{\bf #1},\ #2\ (#3)}
\def\ijqc#1#2#3{{\it Int. J. Quant. Chem.}~{\bf #1},\ #2\ (#3)}

\def\k1{k_1}
\def\k2{k_2}
\def\q1{q_1}
\def\q2{q_2}

\def\({\left (}
\def\){\right )}
\def\[{\left [}
\def\]{\right ]}

\newcommand{\beq}{\begin{equation}}
\newcommand{\eeq}{\end{equation}}

\begin{document}
\date{\today}
\flushbottom \draft
\title{ 
Gaussian process model of 51-dimensional potential energy surface for protonated imidazole dimer
}
\author{Hiroki Sugisawa$^{a,b}$, Tomonori Ida$^b$, and R. V. Krems$^{a,c}$}
\affiliation{
$^a$Department of Chemistry, University of British Columbia, Vancouver, B.C. V6T 1Z1, Canada \\
$^b$Division of Material Chemistry, Graduate School of Natural Science and Technology, Kanazawa University, Kakuma, Kanazawa 920-1192, Japan \\
$^c$Stewart Blusson Quantum Matter Institute, University of British Columbia, Vancouver, BC, Canada V6T 1Z4
}

\begin{abstract}

The goal of the present work is to obtain accurate potential energy surfaces (PES) for high-dimensional molecular systems with a small number of {\it ab initio} calculations in a system-agnostic way. 
We use probabilistic modelling based on Gaussian processes (GPs).
We illustrate that it is possible to build an accurate GP model of a 51-dimensional PES based on  $5000$ randomly distributed  {\it ab initio} calculations with a global accuracy of $< 0.2$ kcal/mol. 
 Our approach uses GP models with composite kernels designed to enhance the Bayesian information content and represents 
the global PES as a sum of a full-dimensional GP and several GP models for molecular fragments of lower dimensionality.
We demonstrate the potency of these algorithms by constructing the global PES for the protonated imidazole dimer, 
a molecular system with $19$ atoms. 
We illustrate that GP models thus constructed can extrapolate the PES from low energies ($< 10,000$ cm$^{-1}$), yielding a PES at high energies ($> 20,000$ cm$^{-1}$). 
This opens the prospect for new applications of GPs, such as mapping out phase transitions by extrapolation or accelerating Bayesian optimization, for high-dimensional physics and chemistry problems with a restricted number of inputs, i.e. for high-dimensional problems where obtaining training data is very difficult.

\end{abstract}

\maketitle

\newpage

\section{Introduction}

Machine learning (ML) is becoming an increasingly powerful tool for
applications in physics and chemistry research.  At the core of these application are models that interpolate in multi-dimensional physical spaces.
These models can be used as surrogates of the solutions of physical equations \cite{gp-book,surrogates,BML}, for optimal control applications \cite{optimal-control}, design, automation and optimization of experiments \cite{ML-for-Chemistry-1,ML-for-Chemistry-2,ML-for-Chemistry-3,ML-for-Chemistry-4} and numerical computations
\cite{ML-for-MD-1, ML-for-MD-2, ML-for-MD-3, ML-for-DFT-1,ML-for-DFT-2, ML-for-DFT-3, ML-for-DFT-4, ML-for-DFT-5, ML-for-DFT-6, ML-for-DFT-7}. 
There are several general approaches to building interpolation models. One is based on parametric models such as neural networks (NN). 
Another is probabilistic modelling, which, in most applications, is based on Gaussian processes (GP) \cite{gp-book}. 
GPs offer several advantages,
including Bayesian algorithms for enhancing model information content \cite{gp-for-PES-4,extrapolation-1,extrapolation-2} and models capable of extrapolation \cite{extrapolation-3,jun-dai}.
The major limitation of GP applications is the numerical difficulty of training and evaluating GP models. 
Training a GP model with $n$ training points
involves iterative inversion of an $n \times n$ matrix, scaling as ${\cal O}(n^3)$, whereas the numerical evaluation of a GP model is a product of two vectors of size $n$, scaling as ${\cal O}(n)$. Therefore, for applications to high-dimensional problems, it is necessary either to introduce approximations that reduce this scaling such as, for example, by data sparsification \cite{sparse-0,sparse-1,sparse-2,sparse-3} or to construct GP models in a way that enhances model accuracy without increasing $n$ \cite{extrapolation-1,extrapolation-2,extrapolation-3,jun-dai}. In the present work, we focus on the latter approach.


A major thrust of recent research has been to develop efficient ML models  for representing potential energy surfaces (PES) for polyatomic systems with accuracy suitable for quantum dynamics simulations \cite{general-fitting-5,ML-for-PES,NNs-for-PES, NNs-for-PESa,NNs-for-PES-1a,NNs-for-PES-1b,NNs-for-PES-1c,NNs-for-PES-2,NNs-for-PES-3,NNs-for-PES-4,NNs-for-PES-5,NNs-for-PES-6,NNs-for-PES-7,NNs-for-PES-8,gp-1,gp-2,gp-3,jie-jpb,gp-for-PES-2,gp-for-PES-3,gp-for-PES-4,gp-for-PES-5,gp-for-PES-6,gp-for-PES-7,gp-for-PES-8,gp-for-PES-9,gp-for-PES-10}.  
 There is also a major effort to develop efficient ML models of force fields for accurate classical dynamics simulations of complex systems \cite{ff-1, ff-2, ff-3, ff-4, ff-5, ff-6, ff-7, carbon-GP, GMDL-1,GMDL-2,GMDL-3,pes-22,pes-33,pes-44,pes-55}. 
This previous work has demonstrated many useful ML approaches to constructing PES and force fields for a variety of systems, including models based on neural networks (NNs) \cite{general-fitting-5, ML-for-PES,NNs-for-PES, NNs-for-PESa,NNs-for-PES-1a,NNs-for-PES-1b,NNs-for-PES-1c,NNs-for-PES-2,NNs-for-PES-3,NNs-for-PES-4,NNs-for-PES-5,NNs-for-PES-6,NNs-for-PES-7,NNs-for-PES-8,pes-22,pes-33,pes-44,pes-55,ff-3,ff-7} and kernel methods \cite{general-fitting-2, gp-1,gp-2,gp-3,jie-jpb,gp-for-PES-2,gp-for-PES-3,gp-for-PES-4,gp-for-PES-5,gp-for-PES-6,gp-for-PES-7,gp-for-PES-8,gp-for-PES-9,gp-for-PES-10,BML,carbon-GP, GMDL-1,GMDL-2,GMDL-3,ff-1,ff-2,rabitz-1,rabitz-2,rabitz-3}, including GP regression \cite{gp-1,gp-2,gp-3,jie-jpb,gp-for-PES-2,gp-for-PES-3,gp-for-PES-4,gp-for-PES-5,gp-for-PES-6,gp-for-PES-7,gp-for-PES-8,gp-for-PES-9,gp-for-PES-10,ff-2}.
 For example, both NNs \cite{NNs-for-PES, NNs-for-PESa,NNs-for-PES-1a,NNs-for-PES-1b,NNs-for-PES-1c,NNs-for-PES-2,NNs-for-PES-3,NNs-for-PES-4,NNs-for-PES-5,NNs-for-PES-6} and GPs \cite{jie-jpb,gp-for-PES-2,gp-for-PES-3,gp-for-PES-4,gp-for-PES-5,gp-for-PES-6,gp-for-PES-7,gp-for-PES-8,gp-for-PES-9,gp-for-PES-10} have been used to produce highly accurate PES for quantum scattering calculations for small systems with $4$ to $6$ atoms.  
GPs with data sparsification have been used to generate high-dimensional force fields for systems as large as bulk crystals \cite{gp-2,carbon-GP}. 
A gradient-domain machine learning (GDML) approach has been recently proposed to obtain force fields for complex molecules by training kernel models
with atomic gradient information instead of energies \cite{GMDL-1,GMDL-2,GMDL-3}. This approach can produce global PES by integrating gradients.  
A significant amount of work has been devoted to building molecular symmetries into the ML models of force fields and PES \cite{gp-for-PES-10,sym-1,sym-2}.

Despite these efforts,
the construction of global PESs with accuracy $\sim 0.1$ kcal/mol for systems with more than 10 atoms remains a challenging task. 
The challenge is due to (i) the complexity of PESs for molecular systems, especially those with multiple different atoms; (ii) the lack of {\it a priori} information on the landscape of PESs, which makes sampling of the configuration space difficult; (iii) the numerical difficulty of high-level {\it ab initio} calculations; (iv)
a wide range of energies spanning both chemical bonds and van-der-Waals interactions that must be simultaneously considered for quantum scattering applications.
 To overcome this challenge, it is important to develop system-agnostic tools for constructing high-dimensional PES that (i) could be applied to different molecular systems, of different dimensionality; (ii) could interpolate and extrapolate quantum chemistry results in order to produce accurate PES using a small number of {\it ab initio} calculations. The ability to extrapolate is essential for the methods to explore the configuration space efficiently using a small number of {\it ab initio} calculations.

The goal of the present work is to obtain accurate PES for high-dimensional molecular systems with a small number of {\it ab initio} calculations $n$ in a system-agnostic way. 
In particular, we aim to obtain GP models of PES with similar accuracy as in previous work on low-dimensional poyatomic systems ($\leq 6$ dimensions), using  similar $n$, but for systems with many more degrees of freedom.
We demonstrate the construction of a 51-dimensional (51D) global PES for a 19-atom system without any information on the evolution of the PES other than a random distribution of potential energy points in a Cartesian space. 
We follow Refs.  \cite{extrapolation-1,extrapolation-2,extrapolation-3,jun-dai}, to improve the interpolation and extrapolation accuracy of GP models in high-dimensional spaces by increasing the complexity of models, instead of increasing $n$, without sparsification.  
We show that this allows us to build GP models capable of interpolation and extrapolation in a 51D space based on $n \approx 5,000$ inputs. 
The present algorithms can be used to model any high-dimensional physics or chemistry problem that depends on a large number ($\sim 50$) of parameters.
This opens up the prospect for applications of GPs, such as non-parametric extrapolation  or acceleration of Bayesian optimization by enhancement of model information content, for high-dimensional physics and chemistry problems with a restricted number of inputs, i.e. for high-dimensional problems where obtaining training data is very difficult.  


\section{Method description}

We begin by a brief description of the conventional algorithm 
for GP regression. A GP $y(\bm x)$ can be considered as a limit of 
a Bayesian neural network with an infinite number of hidden neurons \cite{BML}. In this work, the inputs $\bm x = \left [ x_1, ..., x_N \right ]^\top$ are the variables describing the internal coordinates of a polyatomic system. The output  $y$ is the value of the potential energy.  GPs produce a normal distribution $P(y)$ of values $y$ at any $\bm x$. 
The goal is to condition $P(y)$ by $n$ known values of the potential energy $\bm y = \left [y_1, ..., y_n \right ]^\top$ at $n$ points $\left [ \bm x_1, ..., \bm x_n \right ]^\top$ of the $N$-dimensional variable space. The mean of this conditional distribution at an arbitrary point $\bm x_\ast$ is given by \cite{BML,gp-book}
\begin{eqnarray}
\mu_\ast = \bm k_\ast^\top \bm K^{-1} \bm y,
\label{GP-mean}
\end{eqnarray}
where $\bm k_\ast$ is a vector with $n$ entries $k(\bm x_\ast, \bm x_i)$ and $\bm K$ is a square $n \times n$ matrix with entries $k(\bm x_i, \bm x_j)$. 
The quantities $k(\bm x, \bm x')$ are the kernels, which represent the covariance of the normal distributions of $y$ at $\bm x$ and at $\bm x'$ \cite{BML,gp-book}. Eq. (\ref{GP-mean}) is used to predict the value of the potential energy at  $\bm x = \bm x_\ast$. 

Building a GP model thus reduces to finding optimal kernels $k(\bm x, \bm x')$. To do that, one assumes a simple kernel function, such as, for example, 
\begin{eqnarray}
 k({\bm x}, {{\bm x}'})  =   {\cal M}_v =  \frac{2^{1-v}}{\Gamma(v)}\left( \sqrt{2v}r({\bm x}, {{\bm x}'}) \right )^v \mathcal{K}_v \left ( \sqrt{2v}r({\bm x}, {{\bm x}'})\right )~~~~
\label{eqn:k_MAT}
\end{eqnarray}
where $r^2({\bm x}, {{\bm x}'}) = ({\bm x}- {{\bm x}'})^\top \times {\bm M} \times ({\bm x}-{{\bm x}'})$, ${\bm M}$ is a diagonal matrix with $N$ parameters,  
 $\mathcal{K}_v$ is the modified Bessel function, $\Gamma$ is the
Gamma function, and $v$ is a half-integer. 
The parameters of the kernel function are found by maximizing the logarithm of the marginal likelihood \cite{BML,gp-book}
\begin{eqnarray}
\log {\cal L} = -\frac{1}{2}{\bm y}^\top  {\bm K} ^{-1}{\bm y} - \frac{1}{2}\log |\bm K | - \frac{n}{2} \log 2\pi. 
\label{log-likelihood-explicit}
\end{eqnarray} 

We now make three observations: (i) while Eq. (\ref{GP-mean}) can interpolate any smooth function with any kernel function if $n \rightarrow \infty$, for finite $n$, the interpolation accuracy depends on the functional form of the kernel function $k(\bm x, \bm x')$; (ii) Eq. (\ref{log-likelihood-explicit})  is related to cross entropy of the model and data distributions so maximizing Eq. (\ref{log-likelihood-explicit}) enhances the information content in the model (\ref{GP-mean});  (iii) Eq. (\ref{GP-mean}) uses the training points $\bm y = \left [y_1, ..., y_n \right ]^\top$ directly, so the prediction accuracy is sensitive to the distribution of these points in the $N$-dimensional space.   We exploit these observations to enhance the accuracy of the GP model without increasing $n$.

\begin{figure}[ht]
	\includegraphics[width=1.0\columnwidth]{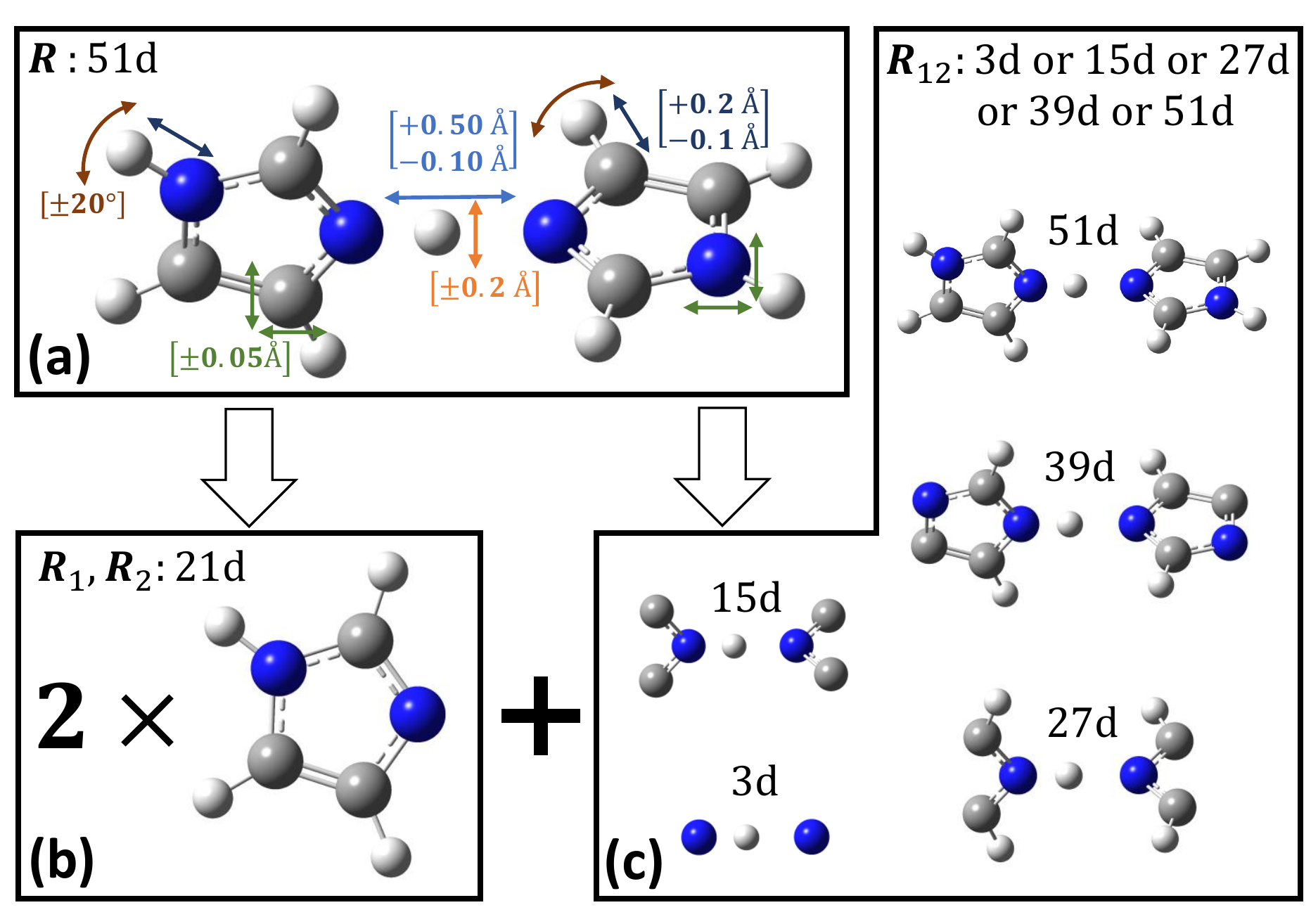}
	\caption{Schematic illustration of the protonated imidazole dimer. Panel (a) shows the coordinate displacements of each atom used to obtain the global surface. Panels (b) and (c) illustrate the fragmentation (\ref{fragmentation}) of the full GP.  
	}
	\label{algorithm}
\end{figure}

The system considered here is the protonated imidazole dimer, shown in  Figure 1 (a). The potential energy of the molecule was calculated using the Gaussian program package \cite{cite-gaussian} at the MP2/6-31++G($d, p$) level of theory. 
To compute the global PES, we started with the known geometry of imidazole dimer in Ref. \cite{previous-geometry}, reoptimized it with the MP2/6-31++G($d, p$) calculations and used the resulting lowest-energy structure as our starting guess. The global deviation of the molecule from this geometry was described using the Cartesian $XYZ$ coordinates for each carbon and nitrogen atom, with the sampling range $[-0.05, +0.05]$ \AA\ for each $XYZ$-direction. 
The coordinate frame was defined by placing the two nitrogen atoms sharing the proton on the $X$ axis, with one of these atoms in the origin of the coordinate frame, and a carbon atom adjacent to the atom at the origin -- in the $XY$ plane.
The configurations for the terminal hydrogen atoms were sampled so that the distance between each hydrogen and its adjacent atom is within $[-0.1, +0.2]$ \AA\ and the angle is $[-20^\circ, +20^\circ]$, as illustrated in Figure 1 (a). 
 Within these coordinate ranges, the potential energy was computed at 15,000 points, randomly generated using the Latin hypercube sampling method to avoid clustering \cite{jie-jpb}. The resulting {\it ab initio} points cover the energy range between zero and $35,000$ cm$^{-1}$. To quantify the accuracy of resulting PES, we compute the root-mean-square error (RMSE) using a large number of {\it ab initio} points that are {\it not used} for training GP models. 

To build the 51D surface, we change the above algorithm for constructing the GP model as follows. 
First, we follow Refs. \cite{extrapolation-1,extrapolation-2,extrapolation-3} to increase the complexity of the kernel function by defining a set of basis kernel functions and combining these basis functions into linear combinations that produce the larger value of ${\cal L}$ in Eq. (\ref{log-likelihood-explicit}). The basis functions include
the functions in Eq. (\ref{eqn:k_MAT}) with $v = 3/2, 5/2$ and $\infty$ as well as the rational quadratic kernel ${\cal M}_{RQ} = \left ( 1 + \frac{|{\bm x}- {{\bm x}'}|^2}{2\alpha\ell^2} \right )^{-\alpha}$.
Note that we use a different metric for model selection from that in Refs. \cite{extrapolation-1,extrapolation-2,extrapolation-3, jun-dai}.  
Second, we follow Refs. \cite{gp-3,molecular-fragmentation} to split the full configuration space into smaller parts and represent the energy of the entire molecular system as
\begin{eqnarray}
E_{\rm total} ({\bm R}) = {\cal E}_{1}({\bm R}_{1}) + {\cal E}_{2}({\bm R}_2) + {\cal E}_{\rm 12}({\bm R}_{12}),
\label{fragmentation}
\end{eqnarray} 
where $\bm R$ is a 51D-vector, ${\cal E}_1$ and ${\cal E}_2$ are independent GP models depending on vectors of lower dimensionality, and ${\cal E}_{12}$ is a GP model that brings the fragments 1 and 2 together into the full surface and that depends on the vector $\bm R_{12}$ with the dimensionality to be determined. 
The model (\ref{fragmentation}) is hereafter referred to as `Composite GP'. 
While the fragmentation (\ref{fragmentation}) is general, here, we use $\bm R_1$ and $\bm R_2$ to represent the separate 21D  imidazole fragments, shown in Figure 1 (b). To determine the dimensionality of $\bm R_{12}$, we construct a series of surfaces, sampling a different number of active degrees of freedom in $\bm R_{12}$, corresponding to the fragments shown in Figure 1 (c). Our results show that $\bm R_{12}$ must account for all 51 dimensions in order for Eq. (\ref{fragmentation}) to be accurate (see Supplementary Material \cite{SM}).


The representation (\ref{fragmentation}) essentially reduces the problem of constructing the 51-dimensional PES to building GP models of potential energy for smaller molecular fragments and constructing a 51-dimensional GP model of the difference between the global surface and these lower-dimensional GPs.
In the following section, we will demonstrate the accuracy gain due to this approach by comparing GP models (\ref{fragmentation}) with those obtained directly by fitting energy in the 51-dimensional space. This approach is motivated by Ref. \cite{molecular-fragmentation} which introduced 
a hierarchy of molecular fragmentations to approximate the total electronic energy from the energies of the fragments. 
It is also analogous to the approach in Ref. \cite{gp-3}, which aims to obtain local energy functionals from total energies. 
In general, the molecular fragmentation for Eq. (\ref{fragmentation}) should be done to ensure that the energy of the fragments and of the full system can be computed using the same {\it ab initio} method.

\section{Results}

The main interpolation results of this work are summarized in Table I, illustrating 
\begin{itemize}
\item that it is possible to construct a 51D surface based on $n = 5,000$  {\it ab initio} energies with the global error under 0.2 kcal/mol; and 
\item how the fragmentation (\ref{fragmentation}) and increasing the
complexity of the kernels improve the accuracy of the resulting surface.
\end{itemize}
 Here, `Simple GP' refers to the 51D model of the surface trained directly by {\it ab initio} points in the $\bm R$-space. 
`Complex $k$' refers to the complex kernels. To identify such kernels, 
we use the greedy-search algorithm -- as in Refs. \cite{extrapolation-3,jun-dai} -- that combines different simple kernel functions in order to maximize the log-likelihood function.  
This algorithm determined the following complex kernels for the composite models:  $k = a {\cal M}_{v = 5/2} + b {\cal M}_{v = 3/2} + c {\cal M}_{v = \infty}$, with $a, b$ and $c$ being free parameters, for ${\cal E}_{\rm 1}$ and ${\cal E}_{\rm 2}$; and  $k = (a {\cal M}_{v = 5/2} 
\times {\cal M}_{RQ}
+ b {\cal M}_{v = \infty} ) \times {\cal M}_{v=1/2} $  for ${\cal E}_{\rm 12}$. 
For the simple GP model with complex $k$, this algorithm determined the kernel $k = a {\cal M}_{v = 5/2} + b {\cal M}_{v = 3/2} + c {\cal M}_{v = \infty} + d {\cal M}_{v = \infty}$ to give the optimal results.  
Note that the parameters of each of the models, including those of the two ${\cal M}_{v = \infty}$ models in the last equation, are independent.
The results labeled `Simple $k$' in Table I refer to GP models with $k = {\cal M}_{v = 5/2}$.

  \begin{table}[t]
  \begin{center}
    \caption{The RMSE for the full 51D surface computed using 10,000 points in the energy range [0, 35000] cm$^{-1}$ as a function of the number of training points $n$.     
    }
    {\tabcolsep = 0.5cm
    \begin{tabular}{crrrr}
      \hline \hline
      Number of & \multicolumn{4}{c}{RMSE {[}kcal/mol{]}}              \\ \cline{2-5} 
      training points& Simple GP & Simple GP & Composite GP & Composite GP \\ 
           &    Simple $k$    & Complex $k$ &    Simple $k$  & Complex $k$ \\ \hline
      1000 & 3.285 & 2.480  & 0.9875 & 0.7837 \\
      2000 & 2.353 & 1.545  & 0.7416 & 0.5373 \\
      3000 & 1.883 & 0.8569 & 0.6203 & 0.4123 \\
      4000 & 1.537 & 0.8666 & 0.5315 & 0.2642 \\
      5000 & 1.286 & 0.7709 & 0.4776 & 0.1815 \\\hline \hline
    \end{tabular}
    }
  \end{center}
  \end{table}

For all of the results in this work, the training energy points are sampled randomly from the indicated energy interval using Latin hypercube sampling to avoid clustering in the configuration space. 
To verify the stability of our results, we performed the following two calculations for the PES obtained with $n=1000$ {\it ab initio} points by interpolation using GPs with complex kernels in the energy range $[0,~35000]$ cm$^{-1}$ (RMSE = 0.7837 kcal/mol as reported in Table I). 
First, we trained a new model of the PES using a different set of 1000 points randomly selected from our set of 15,000 {\it ab initio} points described in the previous section. The RMSE of the resulting surface thus obtained was 0.801 kcal/mol. Second, we computed a new set of 1000 {\it ab initio} points, not included in any of the other training distributions in this paper, and constructed a new PES with these energies as training points. The resulting RMSE was 0.769 kcal/mol. The variation of the RMSE is thus about 2\%. Note that the models using $n=1000$ represent the extreme case and the variation of the RMSE with the randomly selected training distributions must be smaller for models with a larger number of training points.

To illustrate the extrapolation power of high-dimensional GP models, we construct a series of surfaces using the {\it ab initio} points at low energies and predict the global surface at high energies. Table II summarizes the results. The errors reported in Table II are computed using 7,092 {\it ab initio} points in the energy range $[20,000-35,000]$ cm$^{-1}$. Models $A$, $B$ and $C$ are trained by $n$ potential energy points in the energy ranges $[0-35,000]$;  $[0-20,000]$ cm$^{-1}$, and $[0-10,000]$ cm$^{-1}$, respectively. It is impressive to see that model $C$ with $n=5,000$ points, all at energy below $10,000$ cm$^{-1}$, produces a 51D-surface in the energy range $[20,000-35,000]$ cm$^{-1}$ with the global error $\approx 0.6$ kcal/mol. 
This represents the relative average accuracy of better than 1 \% in this energy range. 
   
Models $B$ and $C$ use no information about the PES at energies above $20,000$ cm$^{-1}$. The largest deviation of these model predictions from the {\it ab initio} results in the energy range $[20,000-35,000]$ cm$^{-1}$ is $514$ cm$^{-1}$ for model $A$ and $2740$ cm$^{-1}$ for model $C$ (both for the composite, complex kernel case with $n=5000$). This represents the relative error for that single worst point of $<2.6 \%$ (model $A$) and $< 13.7 \%$ (model $C$). Figure 2 illustrates the accuracy of the interpolation and energy extrapolation of the surface represented by model B with complex kernels.

  \begin{table}[t]
  \begin{center}
    \caption{RMSEs for 7,092 testing points of the energy range $[20,000-35,000]$ cm$^{-1}$ computed for three kinds of models (\ref{fragmentation}) trained by $n$ {\it ab initio} points in 
     the energy range $[0-35,000]$ cm$^{-1}$ (models $A$);  $[0-20,000]$ cm$^{-1}$ (models $B$), and $[0-10,000]$ cm$^{-1}$ (models $C$).}
    {\tabcolsep = 0.5cm
    \begin{tabular}{crrrrrr} 
      \hline \hline
      \multicolumn{1}{c}{} & 
      \multicolumn{2}{c}{Models $A$} & 
      \multicolumn{2}{c}{Models $B$} & 
      \multicolumn{2}{c}{Models $C$} \\
      \cline{2-7}
      $n$& 
      [cm$^{-1}$] & 
      [kcal/mol] & 
      [cm$^{-1}$] &  
      [kcal/mol] & 
      [cm$^{-1}$] & 
      [kcal/mol] \\
      \hline
      1000 & 304.4 & 0.8702 & 332.7 & 0.9513 & 622.8 & 1.781 \\
      2000 & 208.7 & 0.5967 & 220.3 & 0.6299 & 430.0 & 1.230 \\
      3000 & 161.1 & 0.4605 & 168.2 & 0.4810 & 371.3 & 1.062 \\
      4000 & 103.3 & 0.2954 & 115.1 & 0.3291 & 287.9 & 0.8233 \\
      5000 & 71.01 & 0.2030 & 86.53 & 0.2474 & 222.1 & 0.6350 \\ 
      \hline \hline
    \end{tabular}
    }
  \end{center}
  \end{table}

To show that the GP PESs are smooth and physical, we compute the potential energy profile describing proton transfer between the imidazole molecules. Figure 3 shows that the potential energy predicted by the composite GP model (\ref{fragmentation}) trained with $n = 5,000$ {\it ab initio} points is in perfect agreement with the {\it ab initio} results for this minimum energy proton transfer path. Note that the curves shown in Figure 3 represent the minimum of a 51D surface for fixed imidazole - H$^+$ separations.

To illustrate the global performance of the 51D GP PES in the computation of observables, we calculate the vibrational frequencies for the 51 normal modes of the molecule. 
To do that, we diagonalize the Hessian matrix constructed directly from the {\it ab initio} results (hereafter referred to as `Exact') and from the GP models. 
Figure 4 compares the GP model results with the exact results (the numerical values of the vibrational frequencies plotted in this figure are listed in the Supplementary material \cite{SM}). 
Figure 4 shows that all normal modes 
with the frequencies $> 100$ cm$^{-1}$ are well described by the composite GP PES with $n=5000$. Moreover, the GP PES constructed with $n=1000$ {\it ab initio} points captures qualitatively 48 out of 51 normal modes.  This illustrates that a qualitatively correct 51D PES can be constructed with 1000 {\it ab initio} points.

\begin{figure}[ht]
	\includegraphics[width=1.0\columnwidth]{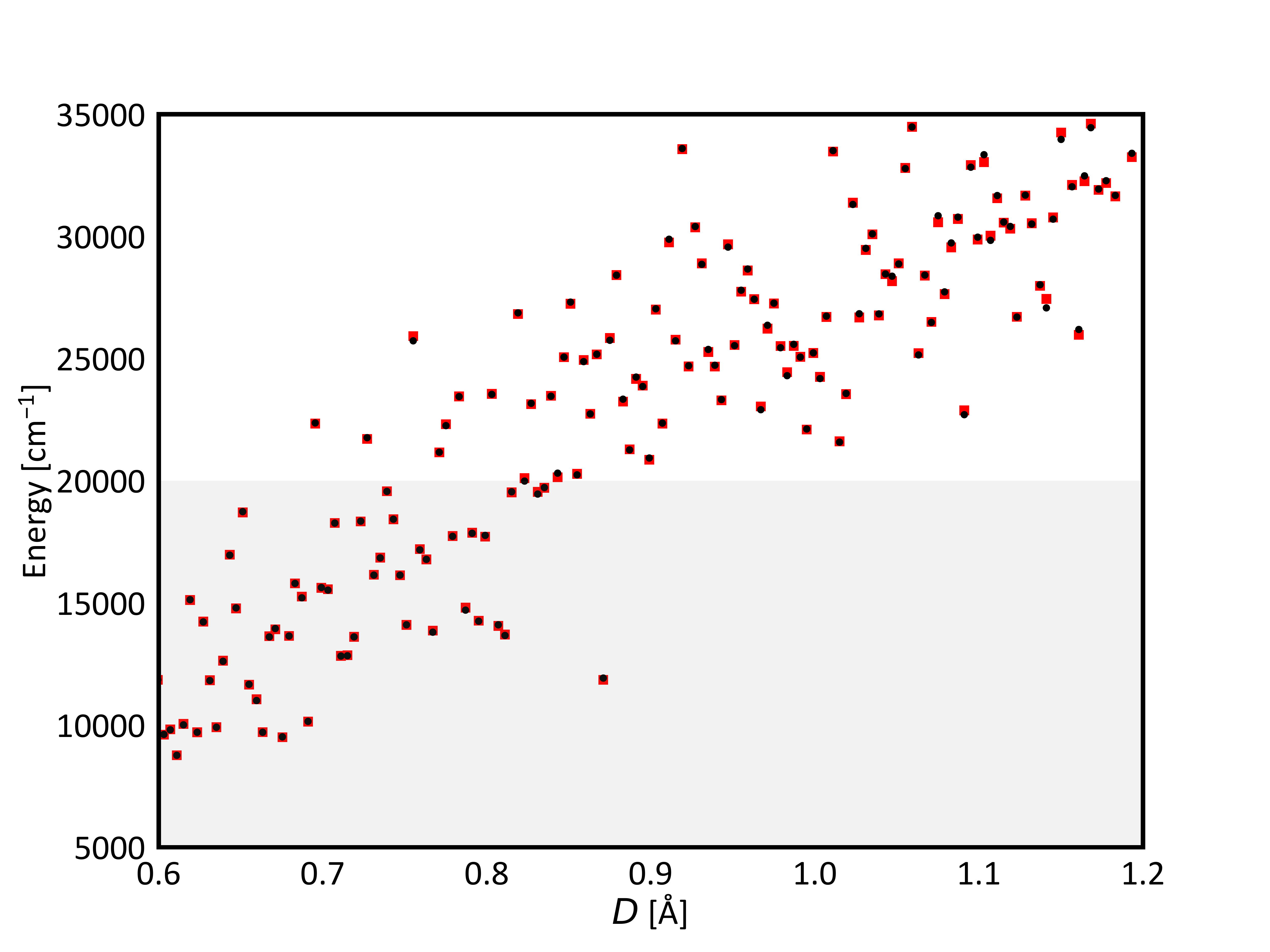}
	\caption{The 51D GP model (circles) in comparison with {\it ab initio} results (squares). The size of the square represents the energy interval 200 cm$^{-1}$. The results are shown for the GP model C trained with 5000 {\it ab initio} points (not shown) at energies below 20,000 cm$^{-1}$ (shaded region).  $D$ is the Euqlidean distance from the equilibrium geometry of the 51D molecule. 
	}
	\label{algorithm}
\end{figure}

\begin{figure}[ht]
	\includegraphics[width=0.8\columnwidth]{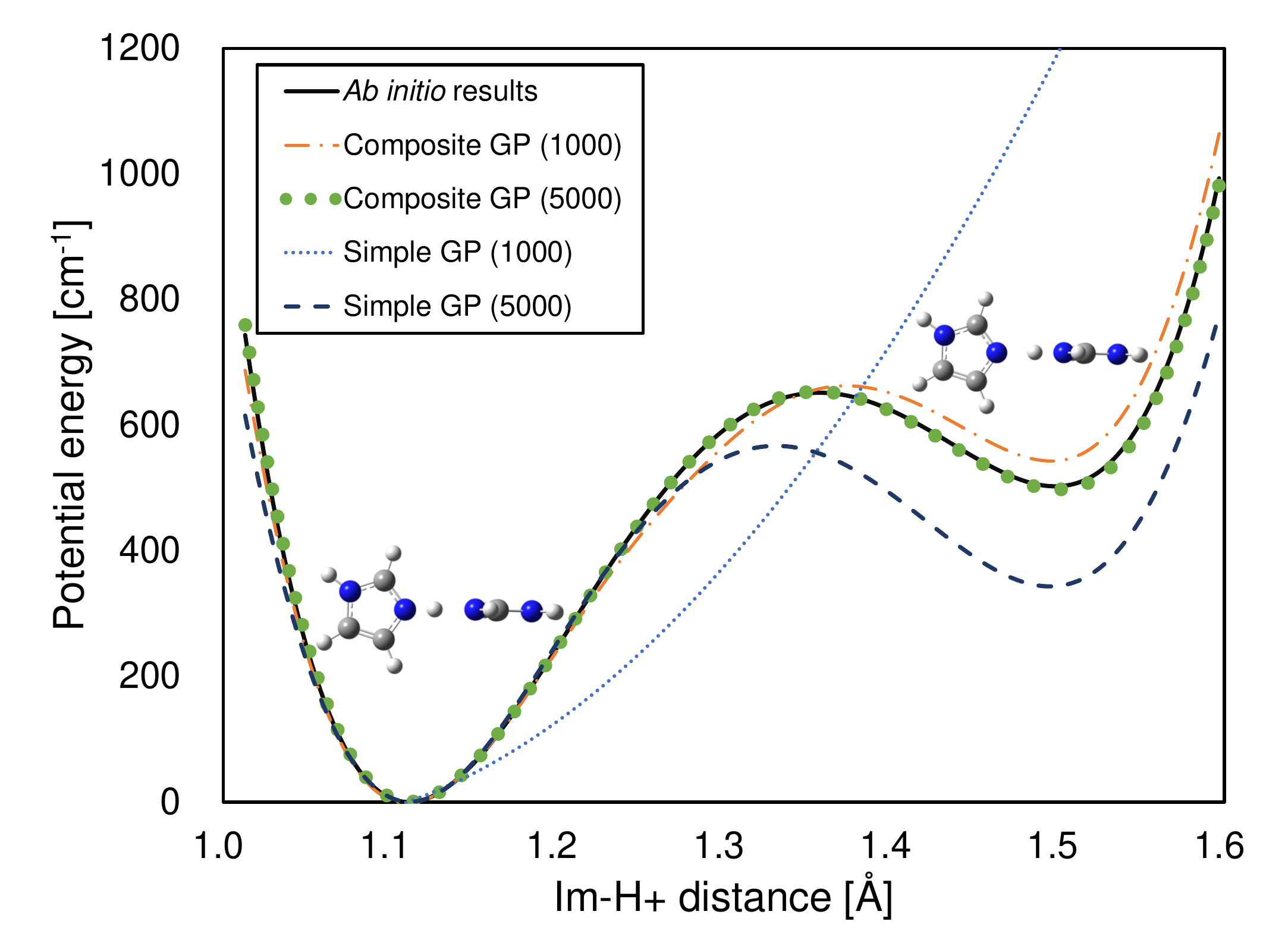} \\
	\includegraphics[width=0.8\columnwidth]{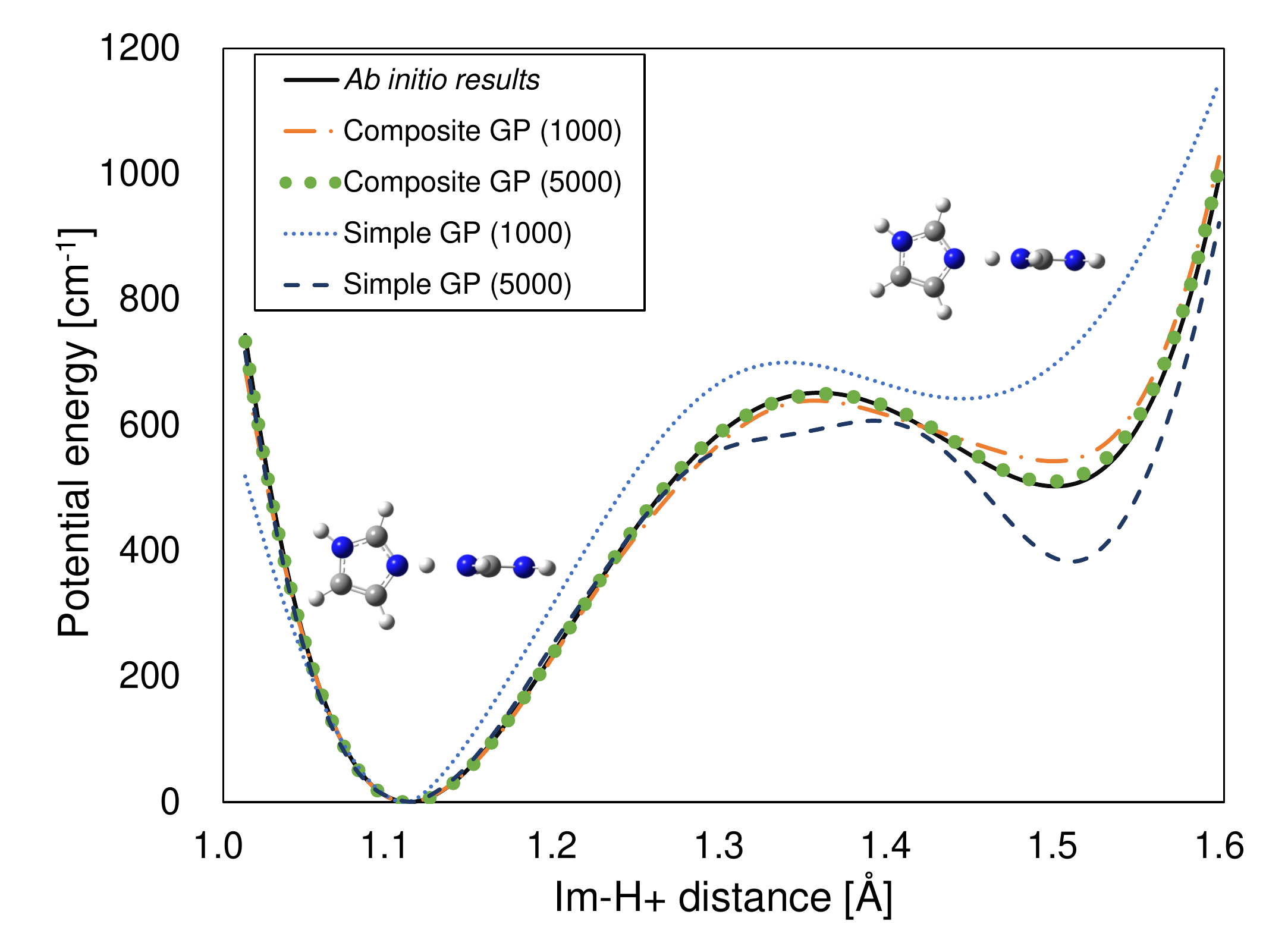} 
	\caption{The minimum energy path for the proton transfer in the protonated imidazole dimer: solid curve - {\it ab initio} calculations; broken curves and green symbols -- the results from the 51D GP models as indicated in the legend box. 
	Upper panel -- results obtained with the simple kernel $k = {\cal M}_{v = 5/2}$; lower panel -- results obtained with complex kernels, as described in text.  
	The green symbols representing the 51D composite GP model are in excellent agreement with the {\it ab initio} calculations. 
	}
	\label{algorithm}
\end{figure}

\begin{figure}[ht]
	\includegraphics[width=1.0\columnwidth]{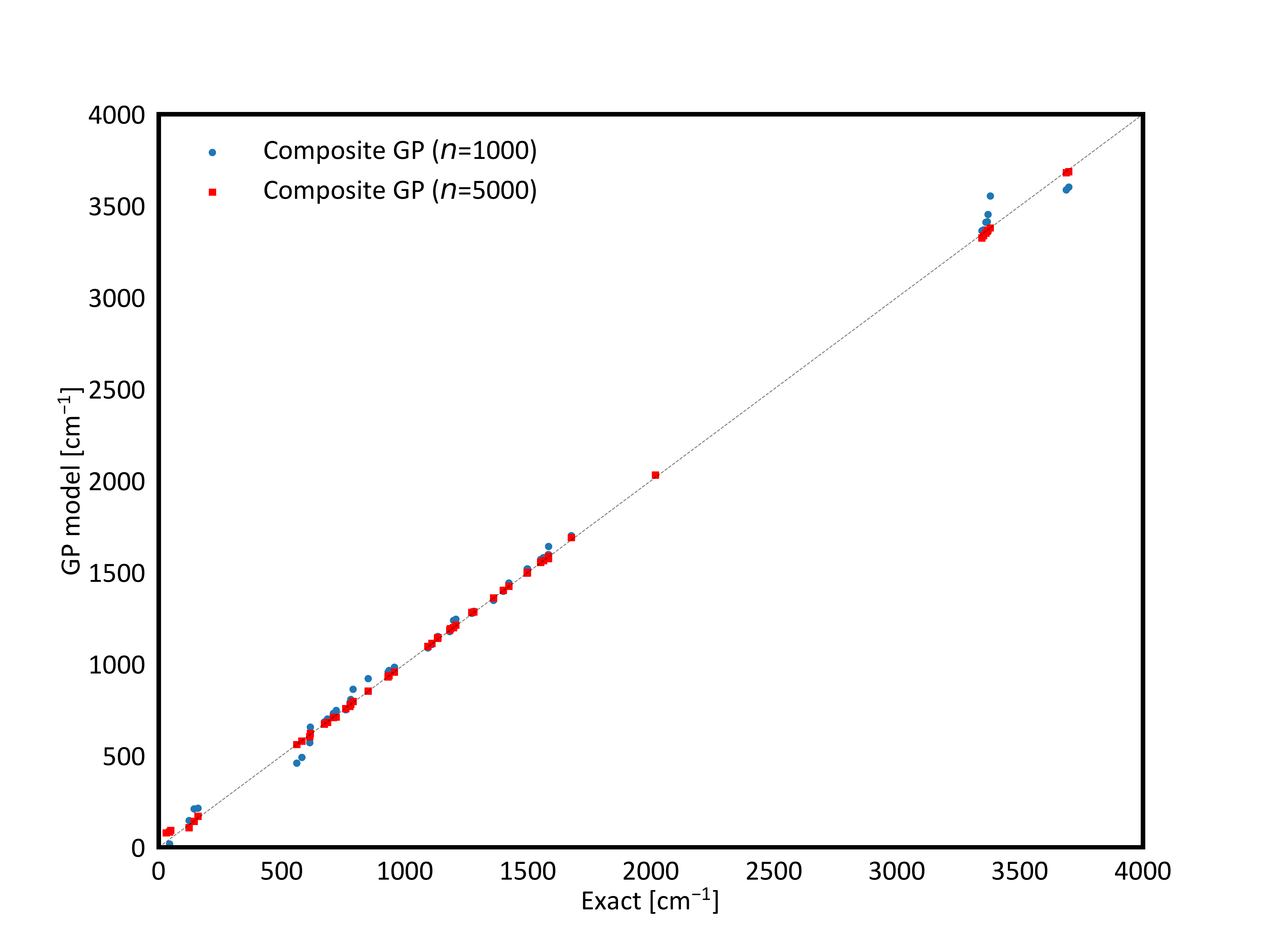}
	\caption{Vibrational frequencies (cm$^{-1}$) for the 51 normal modes of the protonated imidazole dimer computed from the global PES given by Eq. (\ref{fragmentation}) with $n=1000$ (circles) and $n=5000$ (squares). The numerical values of the frequencies are tabulated in the Supplementary Material \cite{SM}. 
	}
	\label{algorithm}
\end{figure}

\clearpage
\newpage

\section{Conclusion}

We have demonstrated an accurate GP model of a 51-dimensional PES for the protonated imidazole dimer (C$_6$N$_4$H$_9^+$)
trained directly by energy points at $5000$ randomly chosen molecular geometries. The PES considered here has a complex landscape, spanning the energy range of 100 kcal/mol. 
It is instructive to compare  the accuracy of the PES obtained here ($0.18$ kcal/mol) with the GDML models in Ref. \cite{GMDL-1} that considered eight molecular systems ranging in complexity from aromatic systems such as benzene with the energy range of 20.2 kcal/mol to aspirin (C$_9$H$_8$O$_4$) with the energy range up to 47 kcal/mol.  
Ref. \cite{GMDL-1} demonstrated that the GDML models trained by 1000 geometries can produce PES with RMSE ranging from 0.09 kcal/mol (for benzene) to 0.36 kcal/mol (for aspirin). The fully converged GDML model for aspirin was shown to produce an RMSE  of about 0.27 kcal/mol. 
We will perform a more direct comparison of these two methods in future work.
We note that the GDML models in Ref. \cite{GMDL-1} use kernel ridge regression with a simple isotropic kernel of the Mat\'{e}rn family. The present work illustrates the accuracy gain due to increasing kernel complexity guided by marginal likelihood maximization. 
 It will be interesting to explore if the accuracy of the fully-converged GDML models can be enhanced by Gaussian process regression (based on marginal likelihood optimization) and by increasing kernel complexity as in the present work.

We note that the accuracy of the models presented here can be further enhanced by increasing $n$ and optimizing the training data distributions. 
Since the form of the kernels in this work is adjusted to the training distributions, an optimal algorithm would require simultaneous optimization of the kernel complexity and the training distributions. The accuracy of the models can also be increased by choosing priors that correspond to the analytic evolution of the PES. However, most of these algorithmic improvements are expected to be system-dependent.

Finally, we have illustrated that 51D GP models with composite kernels can be used to extrapolate PES from low energies ($< 29$ kcal/mol) to high energies ($57$ -- $100$ kcal/mol). 
This opens up the possibility to extend the application of Bayesian methods for searching new physics, such as the approach  in Ref. \cite{extrapolation-3} to identify phase transitions, to high-dimensional physics problems with unknown property landscapes. 
 This can also be used to design efficient methods for Bayesian optimization in high-dimensional spaces \cite{bo1,bo2,rodrigo-bo,BO-highD}. Ref. \cite{BO-highD} illustrated that convergence of Bayesian optimization for two- and five-dimensional problems 
can be accelerated by enhancing GP kernels using Bayesian information criterion for model selection. Since the convergence acceleration is due to the improvement of GP models used for optimization, the present work indicates that a similar acceleration of Bayesian optimization 
should be expected for high-dimensional problems.

\clearpage
\newpage

\section*{Supplementary Material}

Online Supplementary Material presents the numerical values of the RMSE supporting the conclusion regarding the dimensionality of the vector $\bm R_{12}$ in Eq. (4) 
and  the numerical values of the normal mode frequencies depicted in Figure 4. Online Supplementary Material also includes the {\it ab initio} energy points for the protonated imidazole dimer calculated in this work and the python code to construct the 51D PES of the protonated imidazole dimer.

\section*{Data availability}

The data that support the findings of this study are available from the corresponding author upon reasonable request. 

\section*{Acknowledgments}

We thank Rodrigo Vargas-Hernandez for useful discussions and technical assistance. 
HS would like to thank Prof. Mark J. MacLachlan from the University of British Columbia for assistance with the research visit to Canada, during which this work was carried out. 
This work was supported  by TOBITATE! Young Ambassador Program (No. S191N133010001; Japan), JSPS KAKENHI (Grant No. 19K05371; Tokyo, Japan)
and NSERC of Canada.

\widetext
\clearpage

\textbf{\large Supplementary material for `Interpolation and extrapolation in a 51-dimensional variable space: 
system-agnostic construction of high-dimensional PES'}

\vspace{2.cm}

\makeatletter

The purpose of this Supplementary Material is to present the numerical values of the RMSE supporting the conclusion regarding the dimensionality of the vector $\bm R_{12}$ in Eq. (4) of the main manuscript
and  the numerical values of the normal mode frequencies depicted in Figure 4 of the main manuscript. 

\vspace{2.cm}


 
  \begin{table}[h]
  \begin{center}
    \caption{The RMSE of ${\cal E}_{12}(\bm R_{12})$ in Eq. (4) of the main manuscript computed using 10,000 test points and 
     the GP of different dimensionality corresponding to fragments illustrated in Figure 1(c) of the main manuscript. The GP models are trained with 5000 {\it ab initio} points.}
    {\tabcolsep = 0.5cm
    \begin{tabular}{c c } 
      \hline 
Surface dimension & RMSE (kcal/mol) \\      
      \hline 
      
     3d   & 3.045 \\
     15d   & 1.880 \\
     27d   & 1.793 \\
     39d   & 1.276 \\
     51d   & 0.1576 \\ \hline
      
    \end{tabular}
    }
  \end{center}
  \end{table}

    \begin{table}[t]
    \begin{center}
      \caption{Vibrational frequencies (cm$^{-1}$) for the normal modes of the protonated imidazole dimer computed from the global PES given by Eq. (4) with different $n$.}
      {\tabcolsep = 0.5cm
      \tiny
      \begin{tabular}{c c c c c}
        \hline \hline
        \backslashbox[1em]{$n=$}{}  & 1000     & 4000   & 5000   & Exact      \\ \hline
        1  & 3602.3  & 3635.5 & 3686.5 & 3698.6     \\
        2  & 3587.0  & 3624.6 & 3680.9 & 3688.2     \\
        3  & 3553.7  & 3435.0 & 3379.4 & 3379.6     \\
        4  & 3452.8  & 3430.2 & 3367.3 & 3370.3     \\
        5  & 3412.9  & 3380.1 & 3359.9 & 3367.3     \\
        6  & 3410.3  & 3367.2 & 3350.4 & 3361.6     \\
        7  & 3368.5  & 3329.4 & 3337.5 & 3352.0     \\
        8  & 3363.9  & 3323.9 & 3324.8 & 3345.4     \\
        9  & 2031.0  & 2002.5 & 2031.0 & 2018.9     \\
        10 & 1700.1  & 1687.4 & 1689.4 & 1676.8     \\
        11 & 1641.3 & 1584.1 & 1587.0 & 1584.4     \\
        12 & 1597.2  & 1579.3 & 1574.8 & 1583.6     \\
        13 & 1580.6  & 1564.1 & 1563.9 & 1564.0     \\
        14 & 1570.5  & 1551.7 & 1553.9 & 1552.2     \\
        15 & 1519.2  & 1504.6 & 1500.8 & 1498.8     \\
        16 & 1518.7  & 1502.7 & 1495.5 & 1497.3     \\
        17 & 1441.9  & 1426.7 & 1423.7 & 1422.8     \\
        18 & 1396.6  & 1403.0 & 1401.9 & 1400.3     \\
        19 & 1347.2  & 1355.6 & 1360.2 & 1360.6     \\
        20 & 1287.5  & 1277.2 & 1284.0 & 1281.0     \\
        21 & 1276.5  & 1272.7 & 1281.4 & 1272.3     \\
        22 & 1244.3  & 1217.2 & 1212.3 & 1207.6     \\
        23 & 1237.5  & 1203.6 & 1198.2 & 1198.1     \\
        24 & 1190.6  & 1199.3 & 1192.7 & 1185.8     \\
        25 & 1176.9  & 1190.3 & 1185.6 & 1182.7     \\
        26 & 1150.2  & 1131.9 & 1144.0 & 1134.6     \\
        27 & 1139.9  & 1126.4 & 1140.7 & 1132.9     \\
        28 & 1106.2  & 1110.5 & 1112.9 & 1110.3     \\
        29 & 1087.8  & 1088.8 & 1096.3 & 1093.7     \\
        30 & 982.3    & 958.5  & 955.7  & 957.9      \\
        31 & 964.6    & 936.4  & 936.2  & 936.1      \\
        32 & 955.8    & 932.5  & 929.7  & 932.0      \\
        33 & 920.1    & 843.8  & 851.5  & 851.2      \\
        34 & 862.0    & 795.9  & 794.6  & 789.8      \\
        35 & 806.4    & 790.3  & 779.3  & 780.5      \\
        36 & 790.6    & 773.3  & 768.7  & 777.3      \\
        37 & 750.1   & 761.9  & 756.6  & 760.4      \\
        38 & 746.5    & 723.0  & 710.0  & 721.3      \\
        39 & 730.3    & 714.7  & 708.2  & 709.7      \\
        40 & 700.1    & 687.6  & 682.0  & 685.9      \\
        41 & 683.8   & 673.6  & 671.5  & 673.2      \\
        42 & 655.3    & 606.8  & 620.2  & 616.2      \\
        43 & 570.7    & 601.5  & 603.1  & 613.8      \\
        44 & 490.2    & 508.2  & 579.0  & 581.8      \\
        45 & 458.9    & 477.6  & 560.3  & 561.0      \\
        46 & 213.0    & 178.6  & 168.8  & 159.7      \\
        47 & 209.2    & 164.2  & 141.4  & 143.8      \\
        48 & 145.8    & 112.7  & 106.2  & 123.5      \\
        49 & 87.5      & 76.4   & 91.2   & 48.7       \\
        50 & 17.8      & 62.3   & 82.5   & 44.1       \\
        51 & -8.4     & 26.7   & 78.2   & 30.3       \\ \hline \hline

      \end{tabular}
      }
    \end{center}
    \end{table}


\clearpage
\newpage

\begin{thebibliography}{99}

\bibitem{gp-book}
C. E. Rasmussen, and C. K. I. Williams, 
{\it Gaussian Processes for Machine Learning} (The MIT Press, Cambridge, 2006).

\bibitem{surrogates}
J. Cui and R. V. Krems, Gaussian process model for collision dynamics of complex molecules, \prl{115}{073202}{2015}.

\bibitem{BML}
R. V. Krems, Bayesian Machine Learning for Quantum Molecular Dynamics, \pccp{21}{13392}{2019}. 

\bibitem{optimal-control}
M. Benning, E. Celledoni, M. J. Ehrhardt, B. Owren, C.-B. Sch\"{o}nlieb, 
Deep learning as optimal control problems: models and numerical methods, {\it J. Comp. Dyn.} {\bf 6} 171 (2019).


\bibitem{ML-for-Chemistry-1}
R. G\'{o}mez-Bombarelli {\it et al.}, {Design of efficient molecular organic light-emitting diodes by a high-throughput virtual screening and experimental approach}, {\it Nat. Mat.} {\bf 15}, 1120 (2016). 


\bibitem{ML-for-Chemistry-2}
J.N. Wei, D Duvenaud, and A. Aspuru-Guzik, {Neural networks for the prediction of organic chemistry reactions},
{\it ACS Cent. Scie.} {\bf 2}  725 (2016).

\bibitem{ML-for-Chemistry-3}
L. M. Roch, F. H\"ase, C. Kreisbeck, T. Tamayo-Mendoza, L. P.-E. Yunker, J. E Hein, and A. Aspuru-Guzik, 
ChemOS: Orchestrating autonomous experimentation,
{\it Science Robotics} {\bf 3}, 19 (2018).

\bibitem{ML-for-Chemistry-4}
F. H\"ase, L. M. Roch, C. Kreisbeck, and A. Aspuru-Guzik,
Phoenics: A Bayesian optimizer for chemistry,
{\it ACS Cent. Sci.} {\bf 4}, 1134 (2018).


\bibitem{ML-for-MD-1}
P. L. A. Popelier, {QCTFF: On the construction of a novel protein force field}, \ijqc{115}{1005}{2015}.


\bibitem{ML-for-MD-2}
V. Botu and R. Ramprasad, {Adaptive machine learning framework to accelerate ab initio molecular dynamics}, \ijqc{115}{1074}{2015}.



\bibitem{ML-for-MD-3}
M. Caccin, Z. Li, J. R. Kermode, and A. De Vita, {A framework for machine-learning-augmented multiscale atomistic simulations on parallel supercomputers}, \ijqc{115}{1129}{2015}.



\bibitem{ML-for-DFT-1}
J. Wu, Y. Zhou, and X. Xu, {The X1 family of methods that combines B3LYP with neural network corrections for an accurate yet efficient prediction of thermochemistry
}, \ijqc{115}{1021}{2015}.





\bibitem{ML-for-DFT-2}
K. Vu, J. C. Snyder,  L. Li, M. Rupp, B. F. Chen, T. Khelif, K.-R. M\"{u}ller, and K. Burke, {Understanding kernel ridge regression: Common behaviors from simple functions to density functionals
}, \ijqc{115}{1115}{2015}.


\bibitem{ML-for-DFT-3}
J. J. Mortensen, K. Kaasbjerg, S. L. Frederiksen, J. K. N{\o}rskov, J. P. Sethna, and K. W. Jacobsen,
Bayesian Error Estimation in Density-Functional Theory,
{\it Phys. Rev. Lett.} {\bf 95}, 216401 (2005).


\bibitem{ML-for-DFT-4}
A. J. Medford, J. Wellendorff, A. Vojvodic, F. Studt, F. Abild-Pedersen, K. W. Jacobsen, T. Bligaard, and J. K. N{\o}rskov,
Catalysis. Assessing the reliability of calculated catalytic ammonia synthesis rates,
\emph{Science} {\bf 345}, 197 (2014).

\bibitem{ML-for-DFT-5}
M. Fritz, M. Fern\'andez-Serra, and J. M. Soler,
Optimization of an exchange-correlation density functional for water,
\emph{ J. Chem. Phys. } {\bf 144}, 224101 (2016).


\bibitem{ML-for-DFT-6}
R. A. Vargas-Hernandez, Bayesian optimization for tuning and selecting hybrid-density functionals, 
\jpca{124}{4053}{2020}.

\bibitem{ML-for-DFT-7}
J. Proppe and M. Reiher, Reliable Estimation of Prediction Uncertainty for Physicochemical Property Models, 
\jctc{13}{3297}{2017}. 



\bibitem{gp-for-PES-4}
A. Kamath, R. A. Vargas-Hernandez, R. V. Krems, T. Carrington Jr., and S. Manzhos, {Neural networks vs Gaussian process regression for representing potential energy surfaces: A comparative study of fit quality and vibrational spectrum accuracy},
\jcp{148}{241702}{2018}.

\bibitem{extrapolation-1}
D. K.  Duvenaud, H. Nickisch, and C. E. Rasmussen,
{Additive Gaussian Processes}, 
{\it Adv. Neur. Inf. Proc. Sys.} {\bf 24}, 226 (2011).

\bibitem{extrapolation-2}
D. K. Duvenaud, J. Lloyd, R. Grosse, J. B. Tenenbaum, and Z. Ghahramani,
{Structure Discovery in Nonparametric Regression through Compositional Kernel Search},
{\it Proceedings of the 30th International Conference on Machine Learning Research} {\bf 28}, 1166 (2013).


\bibitem{extrapolation-3}
R.Vargas-Hernandez, J. Sous, M. Berciu, and R. V. Krems, 
{Extrapolating quantum observables with machine learning: Inferring multiple phase transitions from properties of a single phase}, 
\prl{121}{255702}{2018}.





\bibitem{jun-dai}
J. Dai and R. V. Krems, 
{Interpolation and extrapolation of global potential energy surfaces for polyatomic systems by Gaussian processes with composite kernels}, 
{\it J. Chem. Theory Comput.} {\bf 16}, 1386 (2020). 


\bibitem{sparse-0}
Y. Cao, M. A. Brubaker, D. J. Fleet, A. Hertzmann, Efficient Optimization for Sparse Gaussian Process Regression, 
IEEE Trans. Patt. Anal. Mach. Intell. {\bf 37}, 2415 (2015) 

\bibitem{sparse-1}
J. Q. Quinonero-Candela and C. E. Rasmussen, A unifying view of sparse approximate Gaussian process regression, {\it J. Mach. Learn. Res.} {\bf 6}, 1939 (2005).

\bibitem{sparse-2}
E. Snelson and Z. Ghahramani, in Advances in Neural
Information Processing Systems 18, edited by Y. Weiss,
B. Sch\"{o}lkopf, and J. Platt (MIT Press, 2006), pp. 1257-1264


\bibitem{sparse-3}
J. Schreiter, D. Nguyen-Tuong, and M. Toussaint, {Efficient sparsification for Gaussian process regression}, {\it Neurocomputing} {\bf 192}, 29 (2016). 












\bibitem{general-fitting-5}
C. M. Handley and P. L. A. Popelier, Potential Energy Surfaces Fitted by Artificial Neural Networks, \jpca{114}{3371}{2010}.


\bibitem{ML-for-PES}
J. Behler, {Perspective: Machine learning potentials for atomistic simulations},
\jcp{145}{170901}{2016}.

\bibitem{NNs-for-PES}
S. Manzhos and T. Carrington, Jr., 
A random-sampling high dimensional model representation neural network for building potential energy surfaces
\jcp{125}{084109}{2006}.


\bibitem{NNs-for-PESa}
S. Manzhos, X. Wang, R. Dawes, and T. Carrington, Jr., 
A nested molecule-independent neural network approach for high-quality potential fits, 
\jpca{110}{5295}{2006}.


\bibitem{NNs-for-PES-1a}
J Behler and M Parrinello, {Generalized neural-network representation of high-dimensional potential-energy surfaces}, 
\prl{98}{146401}{2007}


\bibitem{NNs-for-PES-1b}
J. Behler, {Neural network potential-energy surfaces in chemistry: a tool for large-scale simulations},
\pccp{13}{17930}{2011}.

\bibitem{NNs-for-PES-1c}
J. Behler,  {Constructing high-dimensional neural network potentials: A tutorial review}, \ijqc{115}{1032}{2015}.






\bibitem{NNs-for-PES-2}
E. Pradhan and A. Brown, A ground state potential energy surface for HONO based on a neural network with exponential fitting functions, \pccp{19}{22272}{2017}.

\bibitem{NNs-for-PES-3}
A. Leclerc and T. Carrington, Jr., 
Calculating vibrational spectra with sum of product basis functions without storing full-dimensional vectors or matrices,
\jcp{140}{174111}{2014}.


\bibitem{NNs-for-PES-4}
S. Manzhos, R. Dawes, and T. Carrington, 
Neural network-based approaches for building high dimensional and quantum dynamics-friendly potential energy surfaces,
\ijqc{115}{1012}{2015}.


\bibitem{NNs-for-PES-5}
J. Chen, X. Xu, X. Xu, and D. H. Zhang, A global potential energy surface for the H$_2$ + OH $\leftrightarrow$ H$_2$O + H reaction using neural networks,
\jcp{138}{154301}{2013}.

\bibitem{NNs-for-PES-6}
Q. Liu, X. Zhou, L. Zhou, Y. Zhang, X. Luo, H. Guo, and B. Jiang, {Constructing High-Dimensional Neural Network Potential Energy Surfaces for Gas-Surface Scattering and Reactions},
\jpcc{122}{1761}{2018}.

\bibitem{NNs-for-PES-7}
K. Yao,  J. E. Herr, and  J. Parkhill, 
The many-body expansion combined with neural networks, 
\jcp{146}{014106}{2017}. 


\bibitem{NNs-for-PES-8}
J. Behler and M. Parrinello, Generalized Neural-Network Representation of High-Dimensional Potential-Energy Surfaces, \prl{98}{146401}{2007}. 


\bibitem{gp-1}
C. M. Handley, G. I. Hawe, D. B. Kellab
and P. L. A. Popelier, Optimal construction of a fast and accurate polarisable water potential based on multipole moments trained by machine learning, 
{\it Phys. Chem. Chem. Phys.} {\bf 11}, 6365 (2009).

\bibitem{gp-2}
A. P. Bart\'ok, M. C. Payne, R. Kondor, and G. Cs\'anyi, 
Gaussian Approximation Potentials: The Accuracy of Quantum Mechanics, without the Electrons, 
{\it Phys. Rev. Lett.} {\bf 104}, 136403 (2010). 


\bibitem{gp-3}
A. P. Bart\'ok and G. Cs\'anyi,  
Gaussian approximation potentials: A brief tutorial introduction,
{\it Int. J. Quant. Chem.} {\bf 115}, 1051 (2015).



\bibitem{jie-jpb}
J. Cui and R. V. Krems, {Efficient non-parametric fitting of potential energy surfaces for polyatomic molecules with Gaussian processes}, 
\jpb{49}{224001}{2016}. 

\bibitem{gp-for-PES-2}
P. O. Dral,  A. Owens, S. N. Yurchenko, and W. Thiel, 
{Structure-based sampling and self-correcting machine learning
for accurate calculations of potential energy surfaces
and vibrational levels},
\jcp{146}{244108}{2017}

\bibitem{gp-for-PES-3}
B. Kolb, P. Marshall, B. Zhao, B. Jiang, and Hua Guo, 
{Representing Global Reactive Potential Energy Surfaces Using Gaussian Processes}, 
\jpca{121}{2552}{2017}. 


\bibitem{gp-for-PES-5}
G. Schmitz and  O. Christiansen, {Gaussian process regression to accelerate geometry optimizations relying on numerical differentiation},
\jcp{148}{241704}{2018}.

\bibitem{gp-for-PES-6}
Y. Guan, S. Yang, and D. H. Zhang, {Construction of reactive potential energy surfaces with Gaussian process
regression: active data selection}, \mp{116}{823}{2018}. 

\bibitem{gp-for-PES-7}
G. Laude, D. Calderini, D. P. Tew, and J. O. Richardson, {{\it ab initio} instanton rate theory made efficient using Gaussian process regression}, 
{\it Faraday Discuss.} {\bf 212}, 237 (2018). 


\bibitem{gp-for-PES-8}
Y. Guan, S. Yang, and D. H. Zhang, {Application of Clustering Algorithms to Partitioning Configuration Space in Fitting Reactive Potential Energy Surfaces}, 
\jpca{122}{3140}{2018}.


\bibitem{gp-for-PES-9}
A. E. Wiens, A. V. Copan, H. F. Schaefer, {Multi-Fidelity Gaussian Process Modeling for Chemical Energy Surfaces}, 
{\it Chem. Phys. Lett.} X  {\bf 3}, 100022 (2019). 


\bibitem{gp-for-PES-10}
C. Qu, Q. Yu, B. L. Van Hoozen Jr, J. M. Bowman, and R. A. Vargas-Hernandez, 
Assessing Gaussian Process Regression and Permutationally Invariant Polynomial Approaches To Represent High-Dimensional Potential Energy Surfaces,
\jctc{14}{3381}{2018}.

\bibitem{ff-1}
A. Glielmo, P. Sollich, and A. De Vita, Accurate interatomic force fields via machine learning with covariant kernels, 
\prb{95}{214302}{2017}.

\bibitem{ff-2}
S. T. John and G. Cs\'anyi, Many-Body Coarse-Grained Interactions Using Gaussian Approximation Potentials, \jpcb{121}{ 10934}{2017}. 

\bibitem{ff-3}
K. V. Jovan Jose, N. Artrith, and J. Behler, Construction of high-dimensional neural network potentials using environment-dependent atom pairs, 
\jcp{136}{194111}{2012}.


\bibitem{ff-4}
V. Botu and R. Ramprasad, Learning scheme to predict atomic forces and accelerate materials simulations, \prb{92}{094306}{2015}.

\bibitem{ff-5}
Z. Li, J. R. Kermode, and A. De Vita,
Molecular Dynamics with On-the-Fly Machine Learning of 
Quantum-Mechanical Forces, {\it Phys. Rev. Lett.} {\bf 114}, 096405 (2015).

\bibitem{ff-6}
M. Gastegger, J. Behler, and P. Marquetand,
Machine learning molecular dynamics for the simulation of infrared spectra,
{\it Chem. Sci.} {\bf 8}, 6924 (2017).

\bibitem{ff-7}
L. Zhang, J. Han, H. Wang, R. Car, and W. E,
Deep Potential Molecular Dynamics: A Scalable Model with the Accuracy of 
Quantum Mechanics, {\it Phys. Rev. Lett.} {\bf 120}, 143001 (2018).


\bibitem{carbon-GP}
V. L. Deringer and G. Cs\'anyi, Machine learning based interatomic potential for amorphous carbon, 
\prb{95}{094203}{2017}. 


\bibitem{GMDL-1}
S. Chmiela, A. Tkatchenko, H. E. Sauceda, I. Poltavsky, K. T. Sch\"{u}tt, K.-R. M\"{u}ller, 
Machine learning of accurate energy-conserving molecular force fields,
{\it Sci. Adv.} {\bf 3} (2017). 

\bibitem{GMDL-2}
S. Chmiela, H. E. Sauceda, K.-R. M\"{u}ller, A. Tkatchenko, 
Towards exact molecular dynamics simulations with machine-learned force fields,
{\it Nat. Comm.} {\bf 9}, 3887 (2018). 

\bibitem{GMDL-3}
H.E. Sauceda, S. Chmiela, I. Poltavsky, K.-R. M\"{u}ller, A. Tkatchenko, Molecular Force Fields with Gradient-Domain Machine Learning: Construction and Application to Dynamics of Small Molecules with Coupled Cluster Forces, 2019; arXiv:1901.06594.



\bibitem{pes-22}
J. S. Smith, O. Isayev, and  A. E. Roitberg,
ANI-1: an extensible neural network potential with DFT accuracy 
at force field computational cost, {\it Chem. Sci.} {\bf 8}, 3192 (2017).

\bibitem{pes-33}
K. T. Sch\"{u}tt, F. Arbabzadah, S. Chmiela, K.-R. M\"{u}ller, and A. Tkatchenko
Quantum-chemical insights from deep tensor 
neural networks, {\it Nat. Comm.} {\bf 8}, 13890 (2017).

\bibitem{pes-44}
K. T. Sch\"{u}tt, H. E. Sauceda, P.-J. Kindermans, A. Tkatchenko, and K.-R. M\"{u}ller,
SchNet, A deep learning architecture for molecules and materials,
{\it J. Chem. Phys.} {\bf 148}, 241722 (2018).

\bibitem{pes-55}
J. S. Smith, O. Isayev, and A. E. Roitberg,
ANI-1, A data set of 20 million calculated off-equilibrium conformations 
for organic molecules, {\it Sci. Data} {\bf 4}, 170193 (2017).



\bibitem{general-fitting-2}
T. Hollebeek, T.-S. Ho, and H. Rabitz, Constructing multidimensional molecular potential energy surfaces from {\it ab initio} data, 
{\it Annu. Rev. Phys. Chem.} {\bf 50}, 537 (1999).





\bibitem{rabitz-1}
T. S. Ho and H. Rabitz, 
A general method for constructing multidimensional molecular potential energy surfaces from {\it ab initio} calculations, \jcp{104}{2584}{1996}. 

\bibitem{rabitz-2}
T. Hollebeek, T. S. Ho, and H. Rabitz,
A fast algorithm for evaluating multidimensional potential energy surfaces, \jcp{106}{7223}{1997}. 

\bibitem{rabitz-3}
T. S. Ho and H. Rabitz, Reproducing kernel Hilbert space interpolation methods as a paradigm of high dimensional model representations: Application to multidimensional potential energy surface construction, \jcp{119}{6433}{2003}. 


\bibitem{sym-1}
Y. Guan, H. Guo, D. R. Yarkony, 
Neural network based quasi-diabatic Hamiltonians with symmetry adaptation and a correct description of conical intersections,
\jcp{150}{214101}{2019}.


\bibitem{sym-2}
S. Chmiela, H. E.Sauceda, I. Poltavsky, K-R M\"{u}ller, and A. Tkatchenko, 
sGDML: Constructing accurate and data efficient molecular force fields using machine learning, 
{\it Comp. Phys. Comm.} {\bf 240}, 38 (2019).














\bibitem{cite-gaussian}
Gaussian 16, Revision A.03, M. J. Frisch, G. W. Trucks, H. B. Schlegel, G. E. Scuseria, M. A. Robb, J. R. Cheeseman, G. Scalmani, V. Barone, G. A. Petersson, H. Nakatsuji, X. Li, M. Caricato, A. V. Marenich, J. Bloino, B. G. Janesko, R. Gomperts, B. Mennucci, H. P. Hratchian, J. V. Ortiz, A. F. Izmaylov, J. L. Sonnenberg, D. Williams-Young, F. Ding, F. Lipparini, F. Egidi, J. Goings, B. Peng, A. Petrone, T. Henderson, D. Ranasinghe, V. G. Zakrzewski, J. Gao, N. Rega, G. Zheng, W. Liang, M. Hada, M. Ehara, K. Toyota, R. Fukuda, J. Hasegawa, M. Ishida, T. Nakajima, Y. Honda, O. Kitao, H. Nakai, T. Vreven, K. Throssell, J. A. Montgomery, Jr., J. E. Peralta, F. Ogliaro, M. J. Bearpark, J. J. Heyd, E. N. Brothers, K. N. Kudin, V. N. Staroverov, T. A. Keith, R. Kobayashi, J. Normand, K. Raghavachari, A. P. Rendell, J. C. Burant, S. S. Iyengar, J. Tomasi, M. Cossi, J. M. Millam, M. Klene, C. Adamo, R. Cammi, J. W. Ochterski, R. L. Martin, K. Morokuma, O. Farkas, J. B. Foresman, and D. J. Fox, Gaussian, Inc., Wallingford CT, 2016.

\bibitem{previous-geometry}
G. F. Mangiatordi, J. Hermet, C. Adamo, Modeling proton transfer in imidazole-like dimers: a density functional theory study, \emph{J. Phys. Chem.} A {\bf 115}, 2627 (2011).


\bibitem{molecular-fragmentation}
V. Deev and M. A. Collins, Approximate ab initio energies by systematic molecular fragmentation, 
\jcp{122}{154102}{2005}. 


\bibitem{SM}
The Supplementary Material includes the RMSE values for ${\cal E}(\bm R_{12})$ of different dimensionality and the numerical values of the normal mode frequencies depicted in Figure 4. 



\bibitem{bo1}
J. Snoek, H. Larochelle, and R. P. Adams,
Practical Bayesian optimization of machine learning algorithms,
{\it Adv. Neur. Inf. Process. Sys.} {\bf 25},
{2951}{(2012)}.

\bibitem{bo2}
B. Shahriari, K. Swersky, Z. Wang, R. P. Adams, and N. de Freitas,
Taking the human out of the loop: A review of Bayesian optimization, 
{\it Proc. IEEE} {\bf 104}, 148 (2016).


\bibitem{rodrigo-bo}
R.Vargas-Hernandez, Y. Guan, D. H. Zhang, and R. V. Krems, {Bayesian optimization for the inverse scattering problem in quantum reaction dynamics},  {\it New J. Phys.} (Fast Track Communication) {\bf 21}, 022001 (2019). 


\bibitem{BO-highD}
Z. Deng, I. Tutunnikov, I. Sh. Averbukh, M. Thachuk, R. V. Krems, 
Bayesian optimization for inverse problems in time-dependent quantum dynamics, 
arXiv:2006.06212. 


\end{thebibliography}
\end{document}